\documentclass[conference]{IEEEtran}
\usepackage{cite}
\usepackage{amsmath,amssymb,amsfonts}
\usepackage{graphicx}
\usepackage{textcomp}
\usepackage{xcolor}

\usepackage{siunitx}
\usepackage{multirow}

\usepackage[ruled]{algorithm2e}
\usepackage{algpseudocode}

\def\BibTeX{{\rm B\kern-.05em{\sc i\kern-.025em b}\kern-.08em
    T\kern-.1667em\lower.7ex\hbox{E}\kern-.125emX}}

\begin{document}

\title{Ground Truth Generation Algorithm for Medium-Frequency R-Mode Skywave Detection}

\author{\IEEEauthorblockN{Suhui Jeong} 
\IEEEauthorblockA{\textit{School of Integrated Technology} \\
\textit{Yonsei University}\\
Incheon, Korea \\
ssuhui@yonsei.ac.kr} 
\and
\IEEEauthorblockN{Pyo-Woong Son${}^{*}$} 
\IEEEauthorblockA{\textit{Korea Research Institute of Ships and Ocean Engineering} \\
Daejeon, Korea \\
\textit{Ship and Ocean Engineering Major} \\ \textit{University of Science and Technology} \\
Daejeon, Korea \\
pwson@kriso.re.kr}
{\small${}^{*}$ Corresponding author}
}

\maketitle

\begin{abstract}
With the advancement of transportation vehicles, the importance and utility of navigation systems providing positioning, navigation, and timing (PNT) information have been increasing. 
Global navigation satellite systems (GNSS) are widely used navigation systems, but they are vulnerable to radio frequency interference (RFI), resulting in disruptions of satellite navigation signals. 
Recognizing this limitation, extensive research is being conducted on alternative navigation systems. 
In the maritime industry, ongoing research focuses on a ground-based integrated navigation system called R-Mode. 
R-Mode utilizes medium frequency (MF) differential GNSS (DGNSS) and very high-frequency data exchange system (VDES) signals as ranging signals for positioning and incorporates the existing ground-based navigation system known as enhanced long-range navigation (eLoran). 
However, MF R-Mode, which uses MF DGNSS signals for positioning, exhibits significant performance differences between daytime and nighttime due to skywave interference caused by signals reflecting off the ionosphere. 
In this study, we propose a skywave ground truth generation algorithm that is crucial for studying mitigation methods for MF R-Mode skywave interference. 
Furthermore, we demonstrate the proposed algorithm using field-test data.
\end{abstract}


\section{Introduction}
In the era of the Fourth Industrial Revolution, the significance and applicability of navigation systems providing positioning, navigation, and timing (PNT) information are steadily increasing with the advancements in sophisticated modes of transportation, such as autonomous ships, autonomous vehicles, and urban air mobility (UAM).
This importance is particularly evident in the marine environment, where the absence of surrounding topographical features makes reliable navigation systems crucial.

Currently, the global navigation satellite systems (GNSS) are the most widely used navigation systems \cite{Enge11:Global, DeLorenzo10:WAAS/L5, Chen11:Real, Park20:Effects, Kim18:Low, Lee22:Urban}, utilizing signals transmitted by satellites to provide PNT information. 
The U.S. Global Positioning System (GPS) is a well-known example of GNSS. 
However, GNSS has the disadvantage of being susceptible to radio frequency interference (RFI) \cite{Park21:Single, Park18:Dual, Kim19:Mitigation} due to its weak signal reception strength. 
Additionally, GNSS is vulnerable to ionospheric anomalies \cite{Lee17:Monitoring, Kim14:Comprehensive, Lee22:Optimal, Sun21:Markov, Ahmed17:Statistical}. 
Consequently, there is a growing interest in researching alternative PNT (APNT) systems \cite{Li20, Jia21:Ground, Jeong20:RSS, Han19:Smartphone, Lee20:Integrity, Rhee19:Low, Rhee18:Ground, Kim17:SFOL, Lee22:SFOL, Shin17:Autonomous, Kim17:Simulation, Lee22:Evaluation, Kang21:Indoor, Kim23:Low, Lee23:Performance_Comparison} that can be employed in situations where GNSS is unavailable.

Ranging Mode (R-Mode) \cite{Johnson2014:feasibility, Johnson2014:feasibility1, Johnson2014:feasibility3, Johnson2017:initial, Johnson2020:R-Mode, Son23:Skywave, Swaszek2012:ranging, Jeong21:Development} is an emerging alternative navigation technology in the marine field. 
It offers a terrestrial integrated navigation service that incorporates various components, including medium frequency (MF) R-Mode, very high frequency (VHF) data exchange system (VDES) R-Mode, and the existing terrestrial radio navigation system called enhanced long-range navigation (eLoran) \cite{Son19:Universal, Son18:Novel, Kim22:First, Rhee21:Enhanced, Williams13, Son23:Demonstration, Hwang18:TDOA, Son18:Preliminary, Son19:Preliminary, Son20:eLoran}.

Among these components, the positioning techniques of MF R-Mode and VDES R-Mode utilize the signals of opportunity (SOP) method \cite{Mcellroy2006:navigation, Wang21:Signal, Huang22:Phase, Yang22:UAV}. 
This SOP method leverages the existing radio frequency (RF) signal infrastructure for positioning purposes. 
Specifically, MF R-Mode employs the differential GNSS (DGNSS) signals transmitted through MF bands, which provide GNSS correction information. 
On the other hand, VDES R-Mode utilizes VDES signals for digital communication within the VHF band.

The R-Mode study was initially launched in the EU's R-Mode Baltic project, and the performance of the initial study results was verified in the North Sea area \cite{Johnson2014:feasibility3}. 
The study revealed that MF R-Mode's pseudorange measurement performance was nearly 10 times worse at nighttime compared to daytime. 
The daytime period refers to the time from sunrise to sunset, while the nighttime period refers to the time from sunset to sunrise. 
The primary cause of this phenomenon is the skywave effect on the MF R-Mode signal, which occurs due to signal reflection in the ionosphere during propagation. 
Therefore, research on skywave detection and mitigation is essential to improve the performance of MF R-Mode and minimize its impact. 
However, determining the true occurrence of MF R-Mode skywave, which is crucial for conducting skywave studies, remains unaddressed. Consequently, there is a need to develop techniques and methodologies for generating the ground truth of skywave occurrence in the context of MF R-Mode.

In this study, we propose a skywave ground truth generation algorithm specifically designed for static scenarios, such as between a transmitter and a reference station, to investigate MF R-Mode skywave detection techniques. 
Furthermore, we validate the feasibility of the proposed ground truth generation algorithm using experimental data transmitted from Chungju, South Korea, and received at the Daesan port, South Korea.

\section{MF R-Mode Skywave}
\label{sec:Skywave}

The MF DGNSS signal employed in MF R-Mode is a signal transmitted from a DGNSS reference station. 
It contains GNSS correction information modulated using minimum shift keying (MSK) on a carrier within the 285--325 kHz frequency range. 
However, in MF R-Mode, the DGNSS signal is not used as is. 
Instead, two continuous wave (CW) signals are added to the null frequencies within the DGNSS MSK spectrum.
In the case of the EU's R-Mode Baltic project \cite{Johnson2014:feasibility3}, these frequencies were $\pm$250 Hz  from the center frequency.

\begin{figure}
    \centering
    \includegraphics[width=0.8\linewidth]{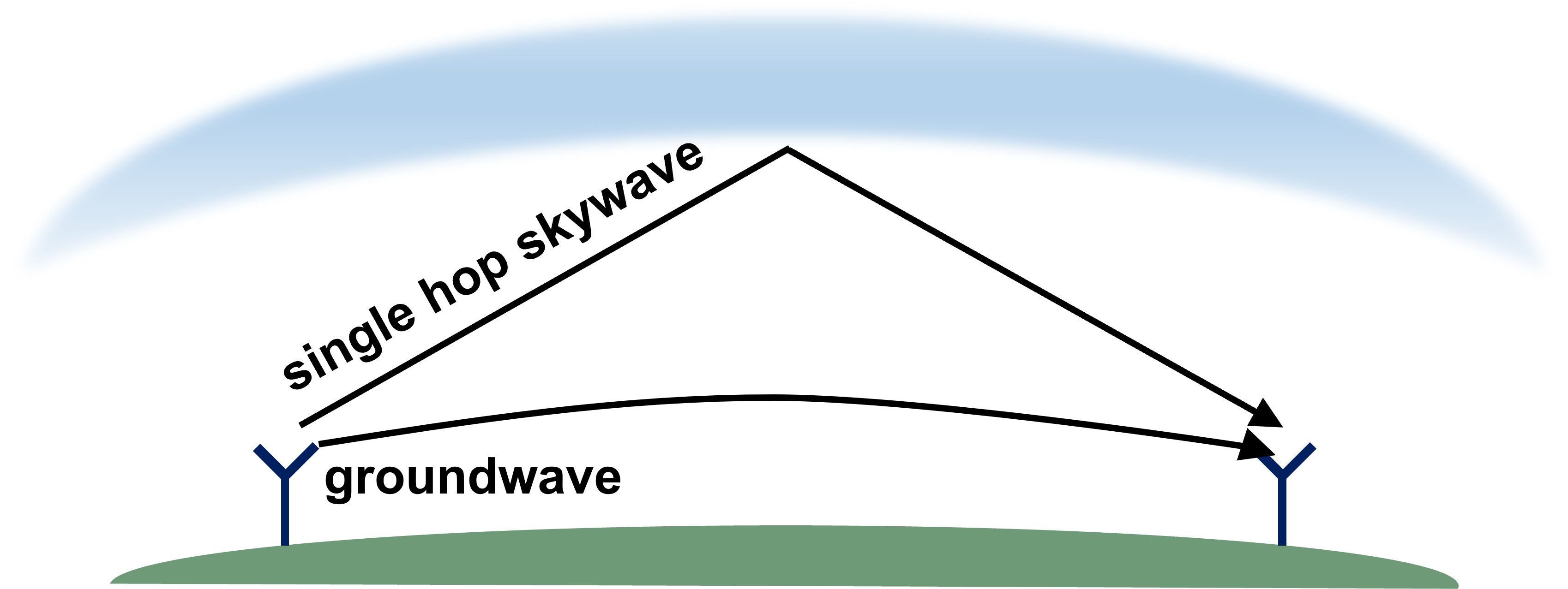}
    \caption{Groundwave and skywave propagation path (modified from Fig. 1 of \cite{Jeong21:Development}).}
    \label{fig:path}
\end{figure}

The generated MF R-Mode signal follows two primary propagation paths from a transmitter to a receiver. 
The first path is known as the groundwave, which is received when the transmitted signal propagates along the Earth's surface. 
The second path is called the skywave, received after reflection from the ionosphere, as shown in Fig. \ref{fig:path}.

During the daytime, primarily the groundwave is received, while during nighttime, both the groundwave and skywave are received, resulting in a significant performance difference between day and night. 
This discrepancy can be attributed to the variations in the ionosphere between daytime and nighttime. 
In the MF frequency range, signals are mostly absorbed by the D layer of the ionosphere, which dissipates after sunset \cite{ITU1998, poppe1995}. 
Consequently, in the evening following sunset, the skywave is generated as signals are reflected by the E layer of the ionosphere. 
Due to the reflection in the ionosphere, the skywave follows a longer propagation path compared to the groundwave, resulting in increased propagation time and attenuation. 
The time delay of the skywave is represented in (\ref{eqn:skywave_timedelay}), where $d$ is the distance between a transmitter and a receiver, $h$ denotes the height of the ionosphere, and $c$ represents the speed of light \cite{Jeong21:Development}.

\begin{equation} 
 \label{eqn:skywave_timedelay}
    t_{d}=\frac{\sqrt{4 h^2 + d^2}-d}{c}.
 \end{equation}
 
The received signal is represented in (\ref{eqn:skywave}), which includes the groundwave along with the skywave reflected once in the ionosphere, known as the single-hop skywave \cite{Johnson2014:feasibility3}.

\begin{equation} 
 \label{eqn:skywave}
 \begin{split}
    r\left ( t \right )
    &= s\left ( t \right )+\alpha s\left ( t-t_{d} \right )\\ 
    &= B\sin \left (  2\pi ft+\phi \right )+\alpha B\sin \left (  2\pi f\left ( t-t_{d}\right)+\phi \right )\\
    &= \eta B\sin \left (  2\pi ft+\phi +\beta \right).
\end{split}
\end{equation}
where $s(t)$ is the groundwave, $s(t-t_{d})$ corresponds to the skywave, $\alpha$ denotes the attenuation factor, $B$ denotes the amplitude of $s(t)$, $f$ is the frequency, $\eta$ is the amplitude scaling, and $\beta$ is the phase shift.
This equation demonstrates that when the skywave and groundwave are received together, the skywave introduces changes in both the amplitude and phase of the groundwave signal \cite{Johnson2014:feasibility3}.

\section{Ground Truth Generation Algorithm}
\label{sec:GroundTruthGenerationAlgorithm}

As mentioned in Section \ref{sec:Skywave}, during daytime, the predominant signal received is the groundwave, while the skywave primarily occurs at nighttime, causing phase and amplitude variations.
In this study, we propose a skywave ground truth generation algorithm specifically designed for static scenarios, such as between a transmitter and a reference station, with a particular focus on the rare occurrence of skywave during daytime.
The algorithm is described in detail as Algorithm \ref{alg:skywave_gt}. 

\begin{algorithm}
\caption{Skywave ground truth generation}
\label{alg:skywave_gt}
\vspace{0.2em}
\KwData{ 1) phase data during daytime for three days, including the day before and after time \textit{t}, $\Phi$ \\
        \hspace*{2.7em} 
        2) phase data at time \textit{t}, $\phi_{t}$}
\KwResult{existence of skywave at time \textit{t}}
\vspace{0.1em}
$\mu _{day} \gets \mathrm{avg}\left( \Phi \right)$ \\
$\sigma _{day} \gets \mathrm{std}\left( \Phi \right)$ \\
$ S \gets \left|\left (\phi _{t}-\mu _{day}\right )/\sigma _{day} \right|$\\
\vspace{0.2em}
\If{\upshape $S \geq 4.5 $} {
        \vspace{0.2em}
        \Return true
        \vspace{0.2em}
}
\Return false
\vspace{0.2em}
\end{algorithm}
Since this is a ground truth generation algorithm, which identifies the actual occurrence of the skywave by post-processing, rather than a real-time skywave detection algorithm, both past and future data with respect to the time of interest can be used. 
To generate the skywave ground truth for the time of interest \textit{t}, the phase data during the daytime is collected for a period of three consecutive days, including the day before and after time \textit{t}. 
The daytime data is obtained from one hour after sunrise to one hour before sunset. 
The mean and standard deviation of the phase data during daytime for the three-day period are calculated. 
Subsequently, using these values, the Z-score of the phase measurement at time \textit{t} is computed. 
Based on the calculated mean and standard deviation of the daytime data, if the absolute value of the Z-score exceeds $4.5$, it indicates the presence of a skywave, representing an outlier in phase variation. 
Conversely, if the absolute value of the Z-score is below $4.5$, it can be inferred that there is no skywave present.

\section{Experiment}

\subsection{Experimental Settings}

\begin{figure}
    \centering
    \includegraphics[width=0.8\linewidth]{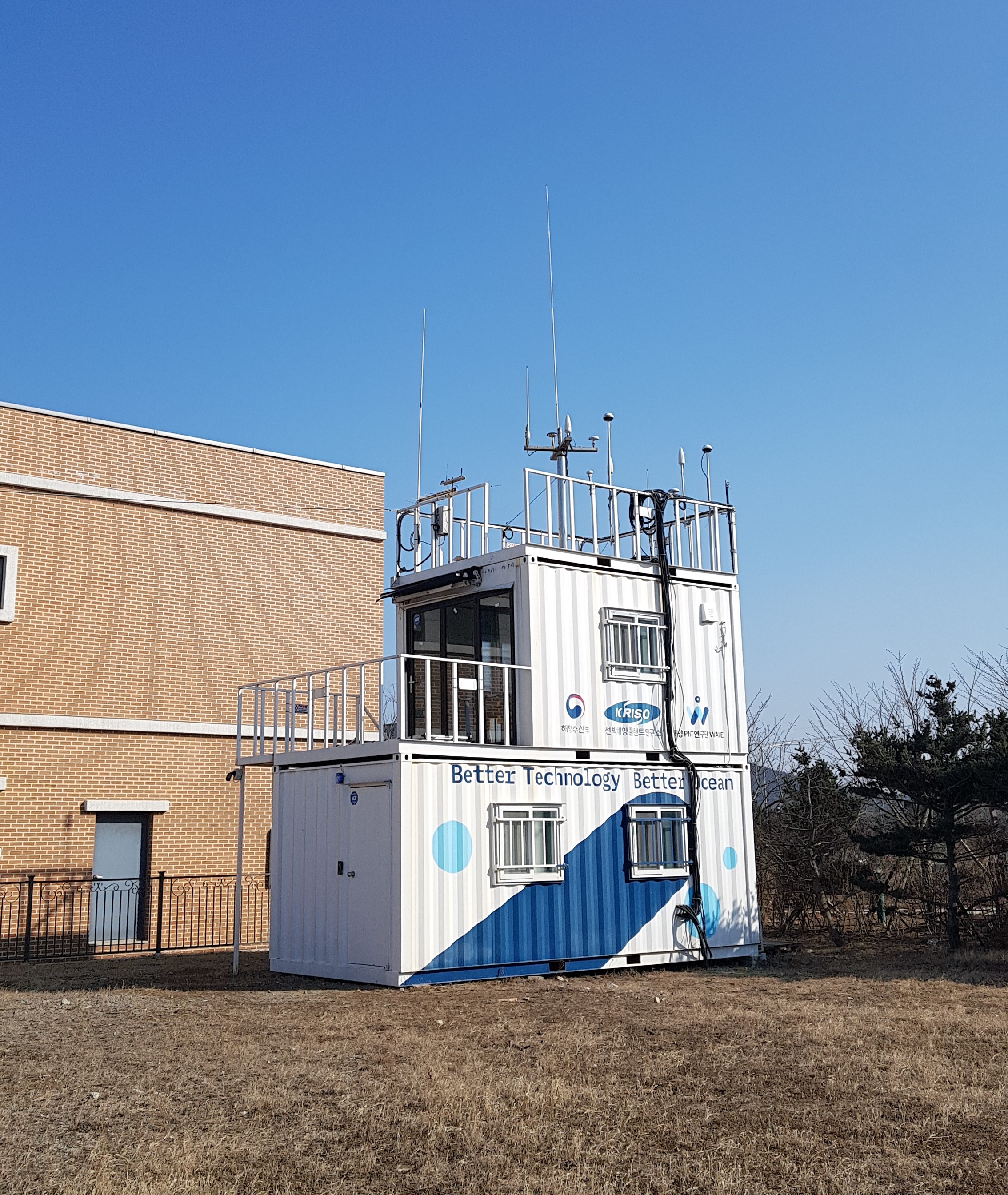}
    \caption{Receiver in the Daesan port.}
    \label{fig:Testbed}
\end{figure}

The experimental verification of the algorithm was conducted by transmitting signals from Chungju, South Korea, and receiving them at the MF R-Mode continuously operating reference station (CORS) located near the Daesan port. 
Fig. \ref{fig:Testbed} illustrates the experimental setup at the CORS in the Daesan port. 
The MSK signal transmitted from the Chungju station has a frequency of 318 kHz, while the CW signals have frequencies centered at 318 kHz $\pm$ 450 Hz.
In contrast to the EU's R-Mode Baltic project, CW signals were added at $\pm$450 Hz from the center frequency to achieve a wider separation between the CW signals.

The receiver used in the experiment is the ``MFR-1a Medium Frequency R-Mode Receiver'' by Serco, which was also used in \cite{Johnson2020:R-Mode}. 
The experiment was performed from February 11th to 13th, 2023. 
Using the proposed Algorithm \ref{alg:skywave_gt}, the ground truth of skywave occurrence in the data on the 12th was computed. 
On February 11th, the sunrise time at the Daesan port was approximately 7:30 AM, and the sunset time was around 6:10 PM. 
For the period from February 11th to 13th, the daytime data used in the experiment spanned from 8:30 AM to 8:00 PM.

\subsection{Experimental Result}

\begin{figure}
    \centering
    \includegraphics[width=0.9\linewidth]{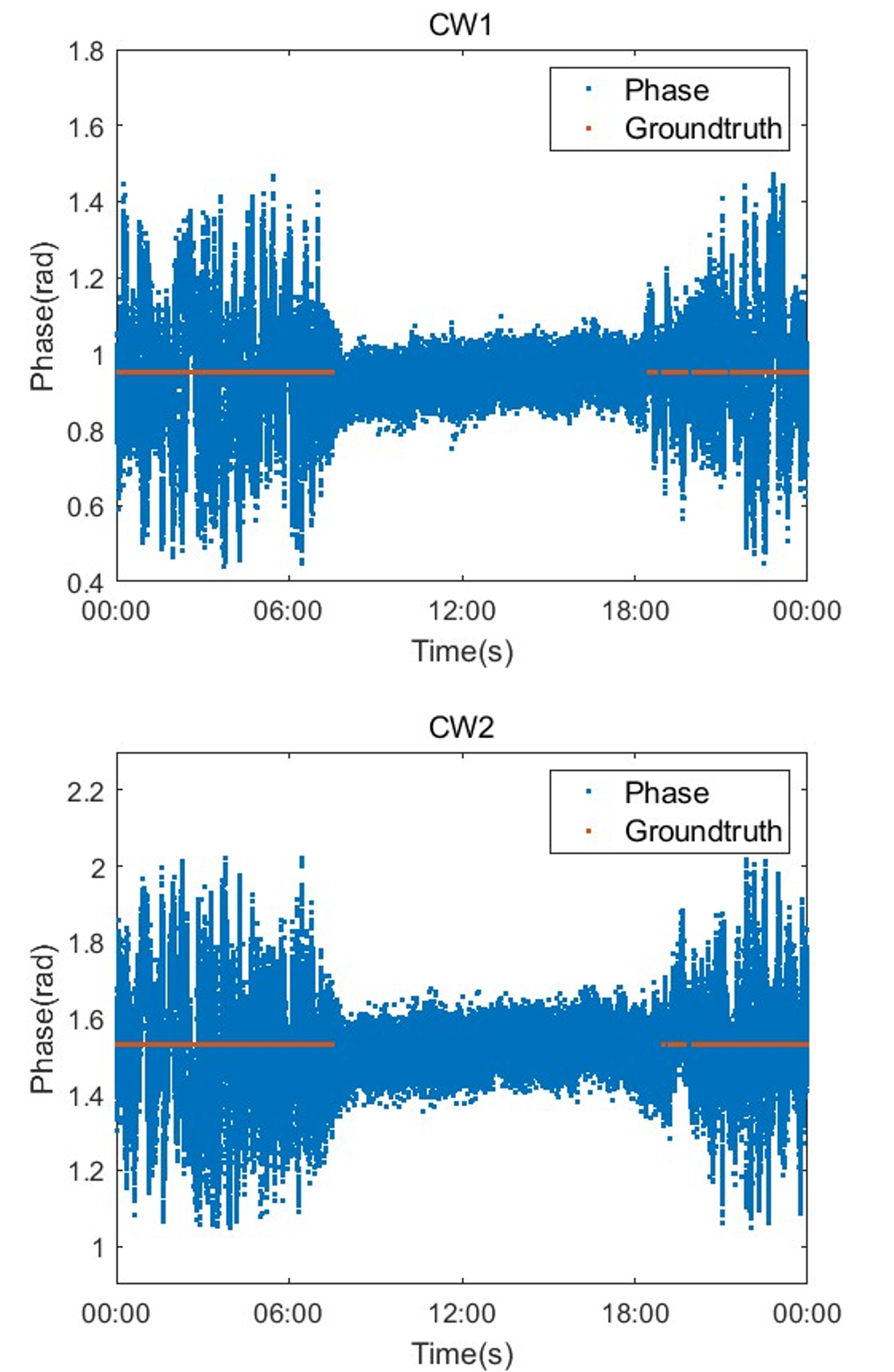}
    \caption{Experimental result.}
    \label{fig:result}
\end{figure}

Fig. \ref{fig:result} shows the results of determining skywave ground truth using the measured phase of the CW1 signal at 317.55 kHz and the CW2 signal at 318.45 kHz at the Daesan port. 
The blue dots represent the measured phase data, while the orange dots indicate the presence of skywave obtained by the ground truth generation algorithm.

\section{Conclusion}

In this paper, we proposed a skywave ground truth generation algorithm that is essential for studying the mitigation of the significant error factor caused by skywave interference in the MF R-Mode system. 
The algorithm was developed based on the observation that skywave phenomena are mainly present during nighttime, accompanied by phase variations, while being largely absent during daytime. 
The validation of the proposed algorithm was performed using MF R-Mode CW1 and CW2 signals transmitted from Chungju, South Korea, and received at the Daesan port.

\section*{Acknowledgment}

This research was conducted as a part of the project titled ``Development of integrated R-Mode navigation system [PMS4440]'' funded by the Ministry of Oceans and Fisheries, Republic of Korea (20200450).
This work was also supported in part by the Future Space Navigation and Satellite Research Center through the National Research Foundation of Korea (NRF) funded by the Ministry of Science and ICT (MSIT), Republic of Korea, under Grant 2022M1A3C2074404; in part by the Unmanned Vehicles Core Technology Research and Development Program through the NRF and the Unmanned Vehicle Advanced Research Center (UVARC) funded by the MSIT, Republic of Korea, under Grant 2020M3C1C1A01086407; and in part by the Korea Institute for Advancement of Technology (KIAT) grant funded by the Korea Government (MOTIE) (P0020535, The Competency Development Program for Industry Specialist).

\bibliographystyle{IEEEtran}
\bibliography{mybibfile, IUS_publications}

\vspace{12pt}

\end{document}